# Neuromorphic Dual-channel Encoding of Luminance and Contrast

Ernest Greene,  Psychophysics Laboratory
Department of Psychology
University of Southern California, Los Angeles, CA 90089
egreene@usc.edu

## Abstract

There is perceptual and physiological evidence that the retina registers and signals luminance and luminance contrast using dual-channel mechanisms.  This process begins in the retina, wherein the luminance of a uniform zone and differentials of luminance in neighboring zones determine the degree of brightness or darkness of the zones.  The neurons that process the information can be classified as "bright" or "dark" channels.  The present paper provides an overview of these retinal mechanisms along with evidence that they provide brightness judgments that are log-linear across roughly seven orders of magnitude.

**Key words:**   neuromorphic   event cameras    retinal encoding    luminance    contrast

*In our technique, if the cell obeys Talbot's Law, drift or fatigue do not affect the measurement, because the output produced by chopped radiation must suffer exactly the same drift or fatigue as that arising with steady radiation of the same average value.  Otherwise the cell would not obey Talbot's law.     F.J.J. Clarke, National Physical Laboratory, U.K.* [1]

"Event cameras" that simulate retinal mechanisms were pioneered by the work of Mahowald & Mead [2] and Zaghoul & Boahen [3] have been proven useful in providing artificial vision to self-driving cars, drones, and autonomous robots. [4]  The are able to resolve thousands of frames per second, with fine temporal resolution, high dynamic range, and high signal-to-noise ratio.[5]  By adding lateral modulation among the pixels, again inspired by retinal physiology, the neuromorphic circuits can perform various image-encoding tasks, including contour detection, image segmentation, registering motion, and classifying texture.[6,7]

Dynamic events in a scene provide the greatest challenge for  machine vision, especially motion and transient fluctuations of light level.  Conceptual insights from psychophysics might prove useful, in particular the ability to specify the perceived brightness of a flickering stimulus as a function of intensity, duration, and frequency. As I will describe below, a long-standing principle called the Talbot-Plateau law can deliver precise predictions about luminance (and brightness) in the presence of fluctuations. As referenced by the quote provided at the outset, an engineering method that is especially precise for encoding light energy was adopted by biological visual a billion of years ago. Circuits that make use of these principles should be more effective in handling the dynamic range of active scenes.

Additionally, retinal physiologists have established dual-channel encoding mechanisms for luminance and contrast. There is evidence that the dual channels provide a linear system for specifying luminance, and encoding of luminance contrast is consistent with the Talbot-Plateau law. Implementing these principles would likely provide discrimination and classification of image content that more closely matched human judgments, which is of special important for anthropic machine vision.

The following sections will describe the Talbot-Plateau principles and provide evidence from my laboratory that tests its validity. Then the retinal physiology that provides for dual coding of luminance and contrast will be described. I will briefly suggest how logic gates might encode the contrast information. However, I will not get into details about engineering options, as those are beyond my level of expertise.

*Calculating Luminance of Flicker-Fused Stimuli*

The flicker stimulus provides a repeating sequence of flashes, with the frequency being sufficiently high that the individual bright and dark portions of each flash cycle are not perceived as separate events. Such a stimulus is described as being "flicker-fused." If the product of frequency and duration of the flicker equals a constant, i.e., the values have reciprocity, one can specify a flash intensity that yield an average luminance that matches the luminance of a steady stimulus. (The relationship actually becomes a three-way reciprocity for frequency by duration by intensity). This principle was first articulated by Joseph Plateau and Henry Talbot, and it was extensively tested and accepted as a "law" in the early part of the 20th century. [8,9] (See Greene & Morrison [10] for a more comprehensive review of the history of perceptual and physiological work on this topic.)

The principle was developed using primitive methods for generating the stimuli and measuring luminance levels, and almost all of the experiments that claimed it to be valid were done using equipment that was not that much better.[11-14] One experiment by Szilagyi and one from my lab used modern electronic equipment and provided some validation, though each was rather limited in the range of treatment combinations.[15,16] However, additional work from my lab has provided strong support for the Talbot-Plateau law.[10]

Figures 1 and 2 show results from one of the experiments reported by Greene & Morrison.[10] Here, respondents were asked to compare the brightness of two stimuli that were successively displayed on an LED array. One of the displays was steady, being varied across five octaves of intensity. Flash sequences were displayed at 24 Hz across each of the treatment combinations, which for these display conditions provided reliable flicker-fused perception.[16] Flash duration was varied across five orders of magnitude, i.e., from 1 to 10,000 microseconds per flash. Flash intensity was varied from trial to trial, providing stimuli that were judged as being darker, brighter, or the same brightness as the steady displays. The data plots shown in Figures 1 and 2 reflect the flash intensities that were seen as matching the brightness of the steady displays.

Figure 1 has plotted the brightness-matching intensities against what would be predicted by the Talbot-Plateau law for eight individual respondents. The group means are shown in the lower right panel. The log-linear precision of the brightness judgments provides substantial support for the validity of the law, at least when considered across the full range of combinations that were provided.

Figure 2 reflects the same data, separated by duration, showing the average intensity of flash sequences against steady intensity. The plot line represents the Talbot-Plateau prediction across the range. Some flash durations departed significantly from the prediction, which could well be due to imperfect calibration. It seems more impressive that the Talbot-Plateau prediction comes very close to specifying the flash intensity needed for a brightness match, even when flash duration was only one microsecond. For a 24 Hz flash sequence, each cycle is over 40 millisecond long (specifically, 41,667 μs), and all of that interval is dark except for the 1-μs flash. Nonetheless, the retina is able to precisely register and signal the average luminance of the flicker-fused stimulus.

These results demonstrated support for the Talbot-Plateau law across roughly seven orders of magnitude. The major relevance of these flicker-fused findings, is that the neural activation by both the bright and the dark portions of the flash sequence must combine to provide luminance that matches the steady luminance. In other words, the light energy being provided by each flash must balance the absence of light during the dark portion of the cycle, and this must match the luminance being provided by the steady stimulus. The retina must register and signal the luminance information from the flicker-fused stimulus with exceptional linearity and precision across a huge range of flash intensity.

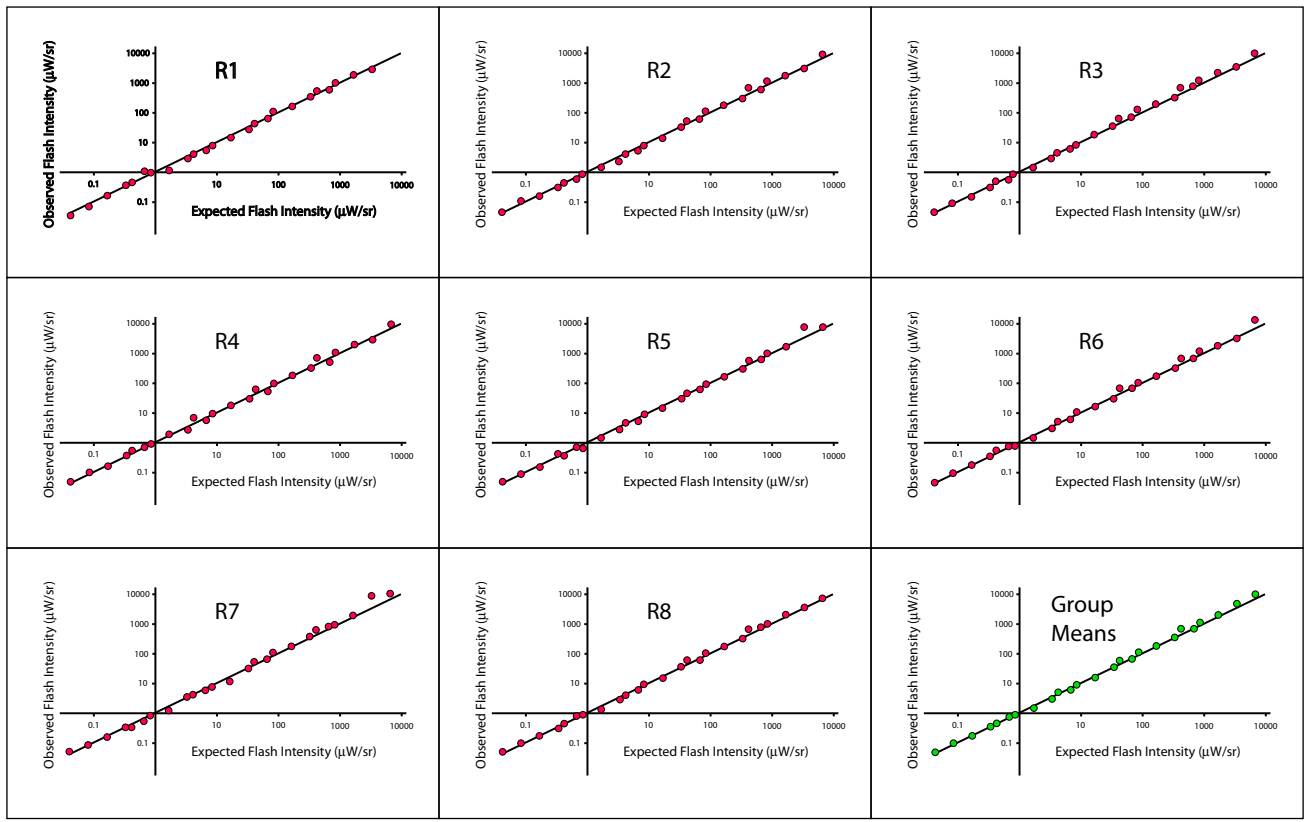

Figure 1. Eight of the panels (R1-R8) show data from individual respondents who compared the brightness of flicker-fused stimuli relative to five octave levels of steady stimuli. All flicker stimuli were at a frequency of 24 Hz, which were consistently perceived as being steady. Flash duration was varied across a range from 1 to 10,000 microseconds. The plot line in each panel shows the Talbot-Plateau prediction of what flash intensity would be required to match the steady intensity, and plot points reflect the observed flash intensity relative to expected flash intensity. The means across the eight respondents are shown in the lower-right panel.

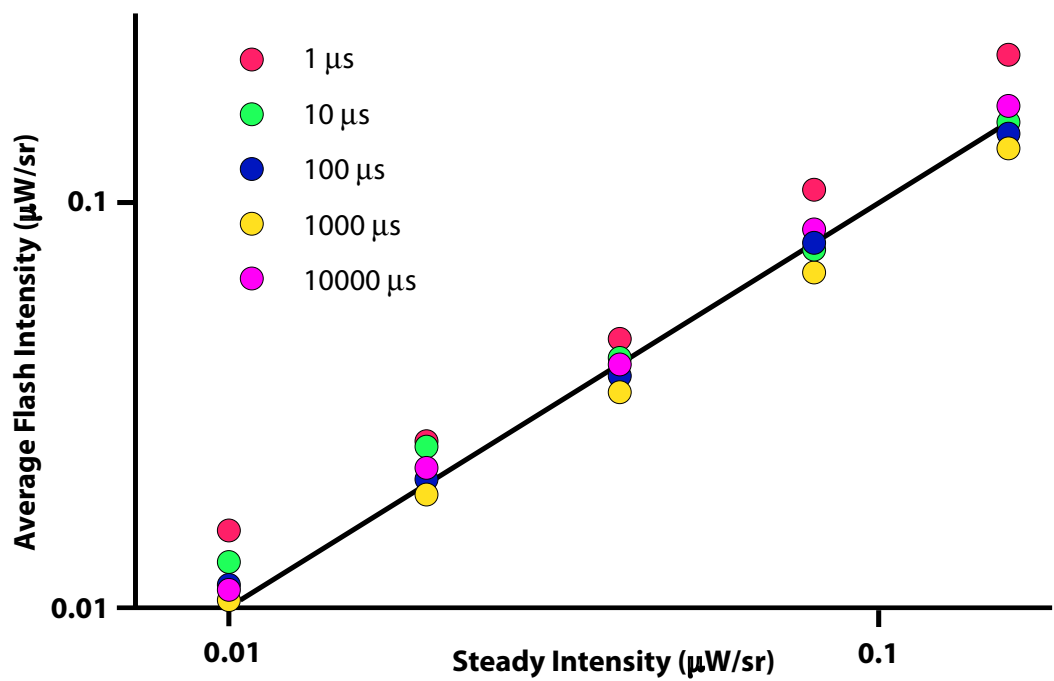

Figure 2.  Here, the mean responses of respondents are re-plotted for each of the flash durations.  The plot line is the Talbot-Plateau prediction for what average flash intensity will match the steady intensity.  Some conditions were found to be significantly different from prediction, but it is unclear whether this was from having reached the limits of linear response of neural mechanisms.

The photoelectric response of silicon is exceptionally linear, but one would not expect the degree of log-linearity reflected in Figures 1 and 2 from a biological system.  Neuron responses to the flash sequences will be discussed below, but here let's note that substantial compensation is often needed to adjust and render neuron signals as linear.  Molnar and associates suggested that interactions between ON and OFF channels provide this benefit, and Werblin cited a number of corrective functions that are accomplished by various sources of inhibition. [17,18]  So one possible reason for encoding luminance using the dual-channel system described below might be to better assure valid measures of luminance.  A dual-coding system might deliver precise log-linear Weber measures by counterbalancing anomalies that one would expect from biological systems.

*Dual-Channel Encoding of Luminance Transitions*

Documentation of that retinal neurons respond differentially to bright and dark light transitions began very early in the last century. [19-23] The photoreceptors register light transitions, which are then passed on to what are commonly called "ON" and "OFF" channels, the main conduits being bipolar and ganglion cells. [24] The way that the retina produces the dual channel is illustrated in Figure 3. The photoreceptor delivers influence to the ON bipolar cell through metabotropic synapses that have the function of reversing its response polarity (Fig 3A), and the synapses connecting to the OFF bipolar cell uses ionotropic synapses that provide no signal reversal. [25-28] Figures 3B and 3C illustrate the direction of change in cone membrane potential as increments and decrements of light level provide selective activation of the ON and OFF bipolar cells. Note that both of the ON and OFF bipolar cells are being depolarized by the respective changes in light level, with an increment of light depolarizing the ON bipolar cell and a decrement of light depolarizing the OFF bipolar cell. These changes in light level are passed on to other retinal units, as will be discussed subsequently.

Labels in Figure 3 have retained the traditional description of bipolar cells as providing ON and OFF responses. While useful for some purposes, it is my view that the terms ON and OFF are not optimal for discussion of luminance signaling or spatial contrast. For one thing, the terms imply transience, where many perceptions are based on continuous signaling from the retina. Further, the terms ON and OFF imply binary states, whereas most of the retinal neurons are registering differential amplitudes of activation. For example, Burkhardt & Fahey reported that ON and OFF bipolar cells in salamander can provide graded biphasic departures from a uniform background level of illumination. [29] Illumination could shift the membrane potential of ON bipolar cells to a higher or lower level. The OFF bipolar cells responded in the opposite direction. The size of the response reflected stimulus intensity, which certainly does not correspond to a binary, fixed amplitude, ON or OFF response. Therefore, going forward, I will use the terms "dark" and "bright" to describe the neuronal activity, with "dark" being used where original reports said "OFF," and "bright" substituting for the term "ON."

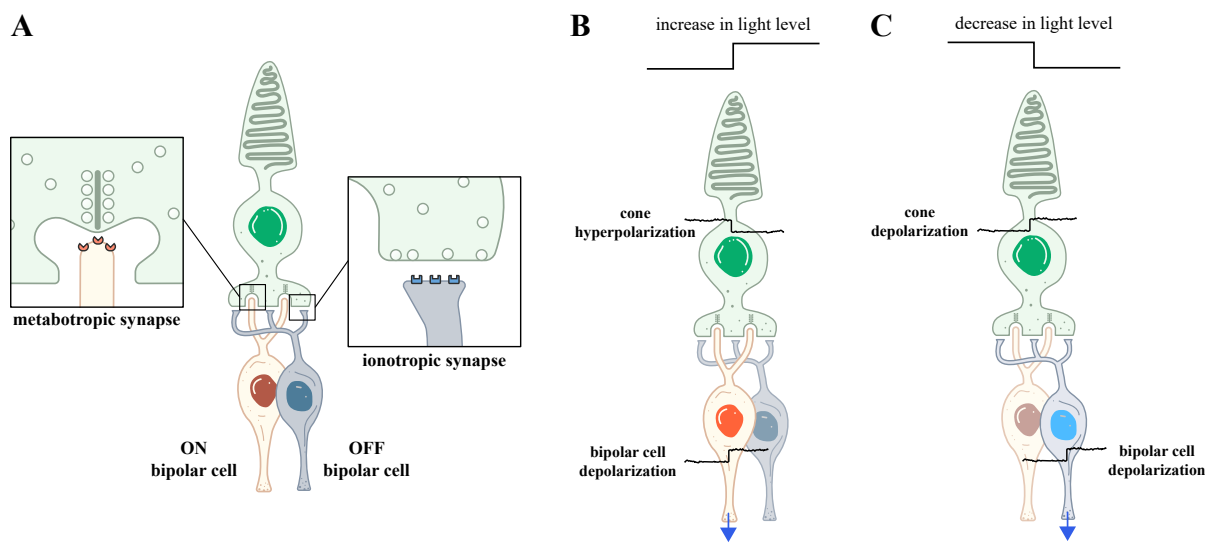

Figure 3. A. The signals provided by retinal cones are first delivered to ON and OFF bipolar cells. Metabotropic synaptic receptors reverse the change cone membrane potential for ON bipolar cells; ionotropic receptors of OFF bipolar cells maintain the same direction of membrane potential change. B and C. Transition of light level bring about hyperpolarizing and depolarizing changes in membrane potentials of cones and bipolar cells. Note that an increment in light elicits a depolarized signal from the ON bipolar cell and a decrement of light elicits a depolarized signal from the OFF bipolar cell.

Below the flicker-fusion threshold the flash sequence produces unambiguous activation of the bright and dark channels. Enroth recorded activity from ground-squirrel ganglion cells and reported seeing discrete spike clusters from each channel at low flicker frequencies.[30,31] These clusters disappeared at about the same frequency at which Dodt found fusion of human ERG responses, which corresponded also to the subjective reports of stimulus fusion.[32] She argued that the flicker-fusion threshold was reached when the ganglion cells could no longer independently register the bright and dark components of the flash sequence. A number of investigators suggested that this occurs as a transition from transient to sustained responding.[33-36] Some said that a dark channel loses its character, i.e., its ability to register dark components of the flash sequence. But spike clusters and attendant oscillations have been recorded in the optic nerve and in other downstream structures at frequencies well above the flicker-fusion threshold.

For example, Laufer & Verzeano found frequency tracking in cats at frequencies in the range of 40-100 Hz. [37]

Further, both channels are contributing to the perception of luminance level. Whether the dark channel is registering each pulse isn't critical, any more than whether the bright channel is registering every flash or some average across flashes. The luminance level is being reported by both channels, one saying how bright it is and the other saying how dark it is. And it does so whether the stimulus is steady or a flicker-fused sequence.

Figure 4 provides a simple diagram of the key elements for signaling luminance levels, these being cones, bipolar cells, and ganglion cells. Rods and their corresponding bipolar cells are not illustrated, as their influence passes through the basic cone channels and they are not essential for present purposes. [38,39] Also absent here are horizontal and amacrine cells, which have lateral communication roles pertaining to center/surround receptive field properties that will be discussed subsequently. The illustration ignores a role for transient responses; it is assuming only tonic activity that signals the luminance levels being registered by cones.

In Figure 4, note that light level of a given zone is being registered and signaled by both bright and dark ganglion cells, with the relative firing rate reflecting the luminance level. Note that dendrites of bright ganglion cells are spread within a strata of bipolar cell telodendria, and are receiving synaptic influence only from the bright bipolar cells. Conversely, the telodendria of dark bipolar cells lie in a different strata, and selectively provide influence on dark ganglion cell dendrites. When there is abundant light stimulation the bright channels are activated more than the dark channels, and with low light it is the reverse. The hypothesis is that this counterbalance compensates for any lack of linear physiological response. The retinal physiology literature is filled with discussion of non-matching bright and dark responses, but these likely pertain to transient operations that other bipolar and ganglion cells are executing. [40]

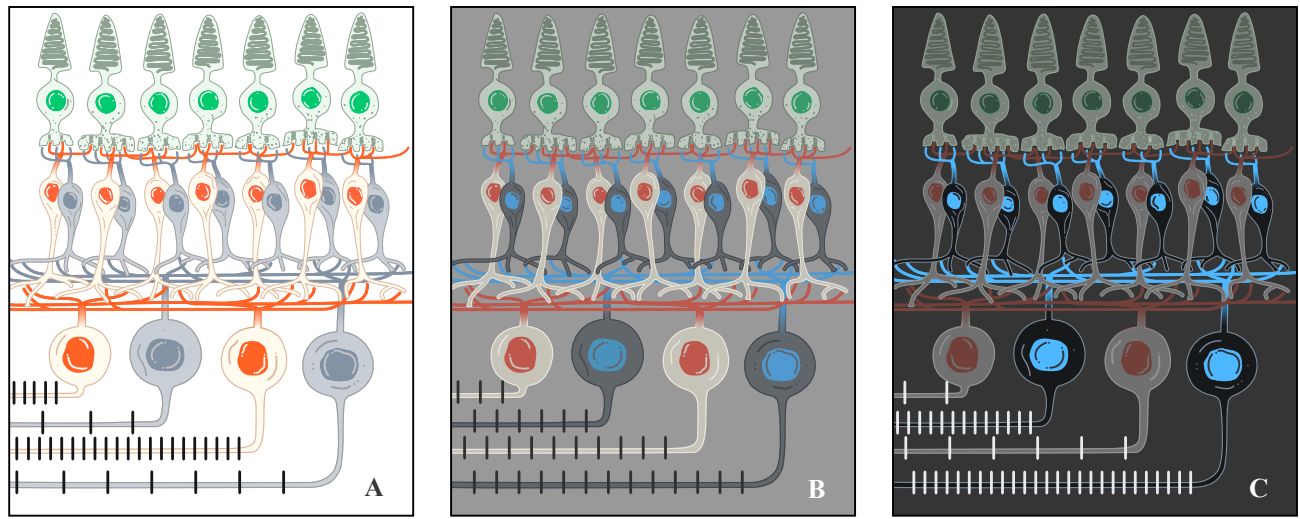

Figure 4. Response of luminance-encoding channels to uniform luminance of a retinal zone. A. A bright stimulus provides strong activation of bright bipolar and ganglion cells, which are marked with bright orange nuclei and dendrites. Dark bipolar and ganglion cells receive weak activation. The spike densities of optic nerve fibers reflect the strong and weak channel activations. B. Medium zone luminance provides roughly equal activation of bright and dark channels. C. A dark zone produces depolarization of cones, which passes directly into the dark channels as strong activation (bright cyan), with the bright channels being weakly activated (dark orange). Across the full range of zone luminance, the bright and dark channels provide for dual-channel encoding of luminance.

The contribution of bright and dark retinal activity is carried forth into the brain, with information being relayed through the lateral geniculate nucleus on the way to primary visual cortex. DeValois and associates recorded differential bright and dark responses from the lateral geniculate nucleus in macaque monkeys.[41] Some of the neurons increased their firing rate as luminance increased, whereas others increased their activity as luminance decreased. Pinneo & Heath described neural activity that was recorded from two patients that had been implanted with depth electrodes in the optic tract and lateral geniculate nucleus.[36] (The electrodes had been implanted for therapeutic purposes, and neither was suffering any visual impairment.) These investigators displayed flash sequences that varied from 1 to 50 Hz, asking the patients to report whether or not they saw flicker or steady stimulation. Net activity from the optic tract of the first patient increased as flash frequency was increased, peaking at 12 Hz, then decreasing with each step up to about 33 Hz, after which the activity level became steady. This was also the frequency at which light emission was perceived to be steady. Brightness of the stimulus remained constant with further increase in frequency. They reported that the second patient gave similar responses. Arduini provided similar findings, noting that above 35 Hz the discharge of bright and dark cannels is constant and comparable to the activity that can be recorded with a gross electrode. [35]

Neurons that respond to steady light levels have been reported in primary visual cortex of monkeys. [42] Firing rate was generally monotonic, either rising or falling as a function of light intensity over a range of about three log units. Many prior investigators had failed to find any units providing steady response that was a function of light level. Doty suggested that this was because of the use of anesthetics and other drugs, which knocked out sustained firing while providing less impairment of transient responses. [43]

Further evidence of dual-channel luminance mechanisms in cortex was provided by Peng & Van Essen. [44] They displayed a large uniform zone that slowly oscillated in luminance across a large range of intensities while recording neuron responses from primary and secondary visual cortex of macaque monkeys. Similar to what Doty and associates (*op. cit*.) had observed, some neurons increased their firing rate as luminance was rising, and others increased their firing as luminance level dropped. Most of the neurons responded to a narrow range of luminance, which suggested to these investigators a neural basis for perception of specific gray levels.

Returning the focus to retinal mechanisms, one would expect that reporting the luminance of a given zone in the image would require a steady retinal signal. However, few investigators have reported sustained (tonic) ganglion cell response to a steady stimulus. Barlow & Levick found just three ganglion cells in cat retina (less than 1% of their sample) that manifested "unusually regular maintained discharge." They called these cells "luxotonic" units. [45]

It is plausible that one needs only a small number of ganglion cells to report the luminance on a patch of retina. There appear to be 30-40 subtypes of ganglion cells that are distinguished by their anatomy and physiology. [46,47] Perhaps each is encoding a specific stimulus attribute, which would include motion-direction selectivity, motion anticipation, and periodic pattern detection. [48-51] (For an overview, see Gollisch & Meister.) [52] One must also consider that a given ganglion cell sub-type may be delivering multiplexed messages which are decomposed by subsequent brain circuits. [53-55]

*Retinal Encoding of Luminance Contrast with Flicker-fused Letters*

In turning to a discussion of luminance contrast, I want to continue to focus on what new lessons might be drawn from the use of flicker-fused displays. Recent work from my lab asked for discrimination of flicker-fused letters against a steady-emission background. [56,57] Letters and background were displayed using an LED array, as illustrated in Figure 5A, with the letters being flickered at various frequencies and intensities using predictions of the Talbot-Plateau law as the basis for specifying their luminance levels. The Talbot-Plateau law correctly specified what combinations of frequency, intensity, and duration of flashes would produce luminance that matched the background luminance. For the combinations at which the luminance levels were balanced, the letters simply disappeared into the background, precluding recognition of the letters (concept illustrated in Fig 5B). Its predictions were fairly accurate for frequencies that ranged from 250 to 250,000 Hz (Fig 5C). This task is essentially a figure/ground contrast discrimination.

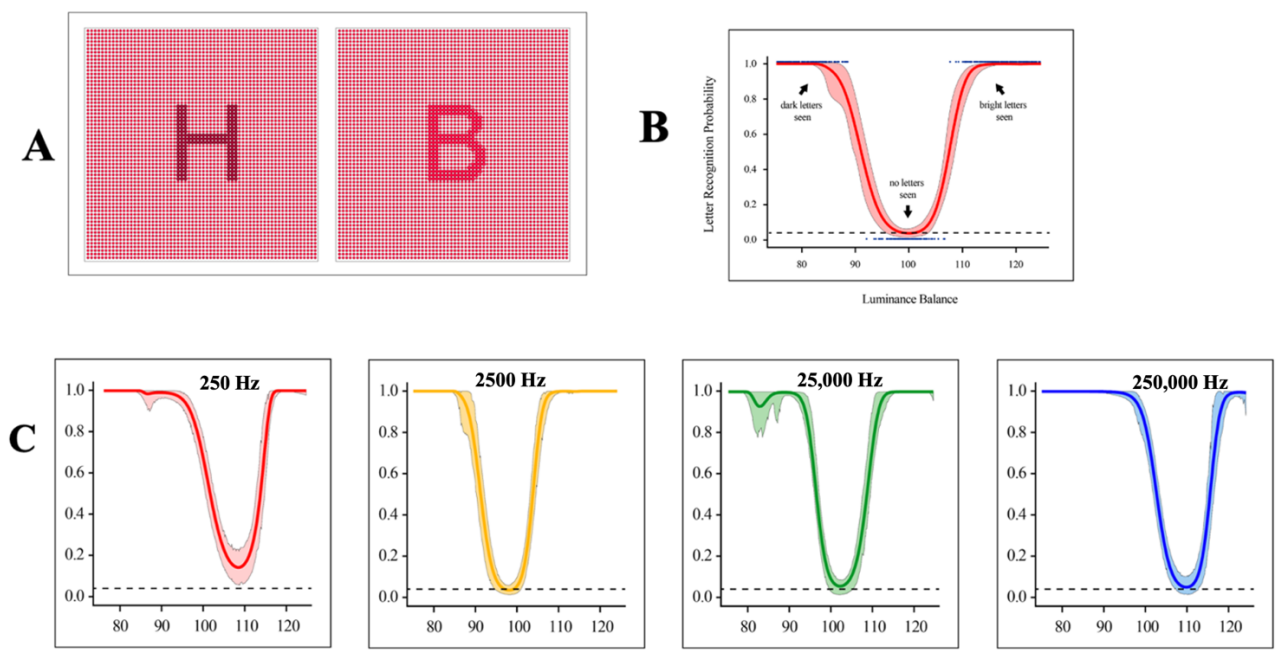

Figure 5. A. Flicker-fused letters can be displayed with average luminance being above or below background luminance. B. Letters can be reliably identified if their average luminance is sufficiently above or below background luminance. They will be invisible if they match background luminance. C. Against the steady background luminance provided in this task, reliable fusion was found at frequencies at or above 50 Hz. Talbot-Plateau's predictions for what treatment combinations would produce matching luminance were supported with flash frequencies being varied from 250 Hz to 250,000 Hz.

Center/surround mechanisms of the retina are especially relevant for discussion of how figures and figure components are discriminated from background. Much of the focus has been on the role of ganglion cells that generate the optic nerve signal. Barlow (frogs) and Kuffler (cats) were the first to note that ganglion-cell activity elicited from a retinal zone could be modulated by stimulating the area that surrounded it. [58,59] The activity relationship matched Hartline's findings of differential spike discharge that could be produced by increments and decrements of light, so the interactions were (and continue to be) described an oppositional ON and OFF neuron activity. [22, 60] Because this is mainly a specification of spatial relationships, I still favor describing the receptive-field activity as interactions among the neurons serving dark and bright channels.

Figure 6A illustrates a luminance differential that is stimulating a bright ganglion cell. The zone that registers the contrast differential is described as the "receptive-field center" of the cell. The higher luminance of this center zone relative to the surrounding area provides the contrast stimulus, which successively activates the bright bipolar and ganglion cells. Conversely, Figure 6B illustrates the activation of dark channels from lower luminance in the receptive-field center. The same zone can serve as the receptive field center for a bright ganglion cell and also be the center of a dark ganglion cell, as the dendrites of each receive connections from bright and dark bipolar telodendria that lie in separate strata (see Fig 4).

The size of the receptive-field center is substantially determined by the area of the ganglion cell dendrites, which can be as small as one cone in the fovea. The full receptive field includes a surrounding retinal zone that can modulate activity of the ganglion cell. A common way of illustrating the center/surround design is provided in Figure 6C. This diagram shows the pool of photoreceptors that can influence a ganglion cell, and implies convergence of influence without showing the intermediate cells that convey the activation. If one is recording action potentials (spikes) with an extracellular electrode, with the retina in darkness or dim illumination, one can record "spontaneous" firing by ganglion cells in the absence of any light stimulus. If one places a narrow beam of light at the center of the receptive field of a bright ganglion cell, one can expect to see an increase in firing rate. Moving that stimulus to the outer edges of the receptive field can suppress firing, and may drop it below the spontaneous rate. The reverse occurs if one is recording from a dark ganglion cell, in that light delivered to the center of the receptive field will suppress firing, and increased firing is seen with stimulation of the surround. Those alternative response outcomes provided the first indication of bright and dark channels. [58,59]

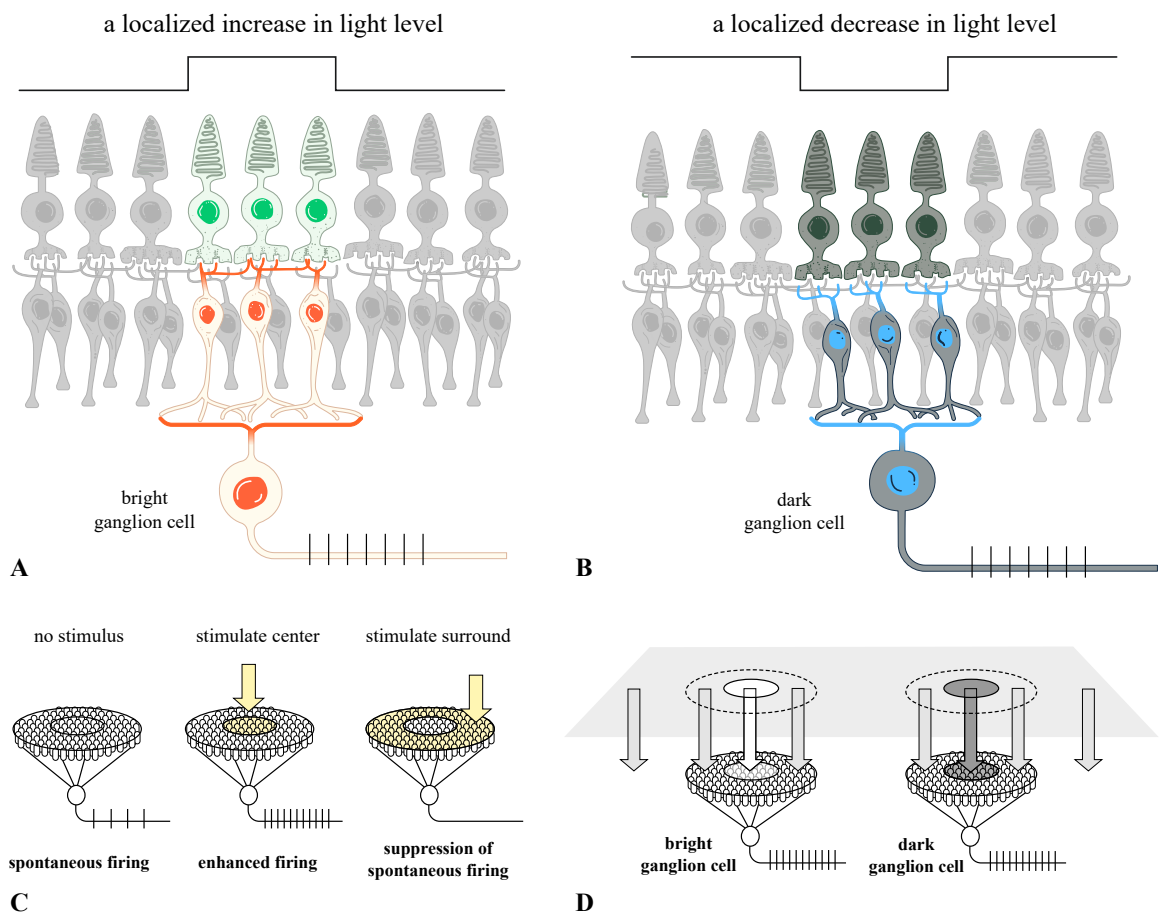

Figure 6. A. Selective center activation of bright ganglion cells is provided by precise synaptic contact of bright ganglion cell dendrites with the end-feet of bright bipolar cells. B. The center of a dark ganglion cell received the same kind of selective activation from dark bipolar cells. C. This is a classic method of illustrating that most ganglion cells receive influence from a surrounding zone that will modulate (inhibit) activity being delivered from the center. It diagrams how light provided to the center activates the ganglion cell, and light provided to the surround suppresses activity. D. A better way to illustrate the center/surround concept shows the bright and dark stimuli that will activate the respective ganglion cells, with those stimuli being surrounded by background. This makes it clear that the center/surround mechanism is designed to register the contrast of figures and figure elements relative to background.

I think the Figure 6C illustration is misleading. The surround is shown as a ring (an annulus), with light stimulation of that ring modifying the ganglion cell's firing rate. The illustration is almost implying that the ganglion cell is designed to register light-rings. I favor the diagram provided in Figure 6D. This shows how the bright and dark ganglion cells would respond to a difference in light level falling only on the center, with the surround being stimulated by a uniform background. If the luminance level being provided to the center is higher than background, the bright ganglion cell will be activated. A spot providing a luminance level that is less than background will activate a dark ganglion cell. The illustration shows the two ganglion-cell types at separate locations, but one will recall from Figures 6A & B that bright and dark channels are registering the same pool of photoreceptors. So if one were to shift the dark ganglion cell in Figure 6D to the left, placing its receptive field center coincident with the center of the bright ganglion cells, the pair would register any changes in light level at that common center. The dual-channel design signals contrast of a common center zone relative to luminance of the surrounding area.

Figure 6D better conveys the concept that spatial contrast is primarily a figure/ground discrimination. However, it is still misleading with respect to the scale of the surround influence. Flores-Herr and associates assessed dimensions of center and surround of ganglion cells in rabbits. [61] The size of the centers matched the dimensions of the ganglion cell dendrites, but diameters of the surround influence were several times larger. Measures on upstream neurons show similar size differences. Surround influence on macaque bipolar cells is 2-3 times larger than the diameters of the bipolar dendrites. [62] Similarly, Verweij and associates assessed the size of the surround annulus that would suppress activity of single cones in macaque. [63] The center response of the cone had a span of about 23 microns, and mean diameter of the surround annulus was 451 microns. As discussed below, the surround influence is thought to be provided by horizontal cells. The span of horizontal-cell dendrites is very large and they are known to link together to provide a functional syncytium as a function of light level. At the eccentricity that was tested, the span of the surround effect was larger than the diameter of horizontal dendrites, which may reflect coupling of the horizontal cells. Similar findings were reported by Packer & Dacey. [64] Whether a single horizontal cell or several, the size of the surround is certainly much larger than suggested by most diagrams, including those in Fig 7C and 7D.

*Contrast Modulation by Horizontal Cells*

It is now well established that horizontal cells provide modulation of cone and bipolar cell activity, and are the major source of surround influence. These cells have wide dendritic arbors that make intimate contact with cones and bipolar cells, as illustrated in Figure 7. Excellent reviews of what is presently known about these physiological and pharmacological mechanisms has been provided by others and will not be repeated here. I mainly want to note what may be pertinent to the encoding and signaling of flicker-fused stimuli. Since the use of flicker has not commonly been used in the study of retinal mechanisms, much of the following will be speculative, with considerable uncertainty about what conclusions should be drawn.

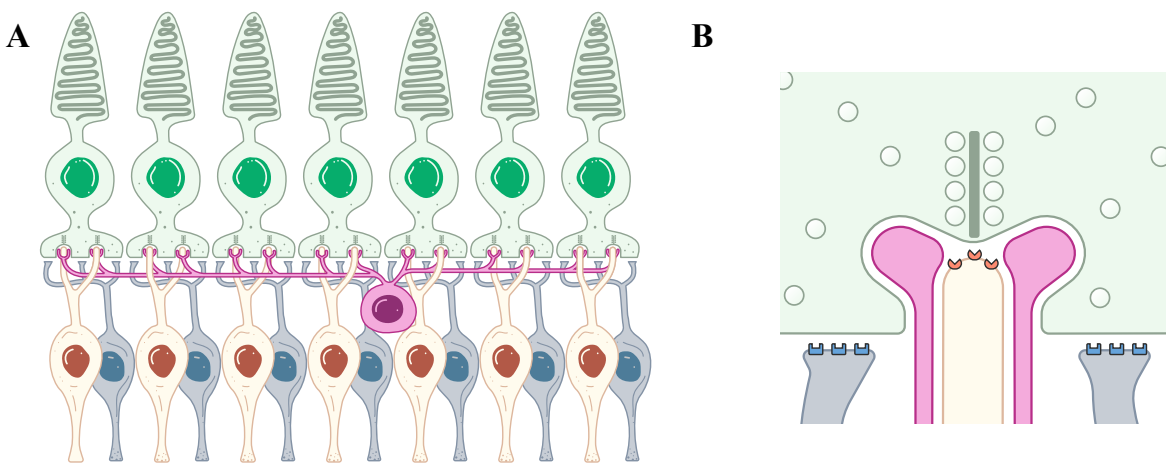

Figure 7. A. Horizontal dendrites spread laterally to receive cone activation and provide feedback. B. Horizontal dendrites occupy the cone ribbon synapses, providing modulation of both bright and dark bipolar cells. [28]

Werblin & Dowling provided the earliest evidence that horizontal cells provide surround influence. [65] Intracellular recordings from salamanders established the central zone that would activate either dark- or bright- responding bipolar cells, and said that the dimensions of the center matched the spread of bipolar-cell dendrites. Antagonism of that activity was produced by an annulus of light that activated the horizontal cell. The effects of center stimulation always preceded the influence of peripheral stimulation, which implied a synaptic basis for the modulating effects. Dacheau & Miller found similar results with rabbit. [66]

Werblin & Dowling did not observe any sign of surround influence on photoreceptors. [65] Perhaps these were rods that generally have little or no center/surround interactions. Baylor and associates did find surround influence on cones that they attributed to horizontal feedback. [67] Other investigators have supported the Baylor report. [68-70]

Based on these and related findings, there is substantial consensus that the surround influence that can be observed in downstream cell populations begins with horizontal-cell feedback onto photoreceptors and bipolar cells. There are, however, a number of ambiguities with respect to this mechanism. In my view, the entire center/surround model is poorly framed. The influence is most often described as "inhibition," which does not adequately convey the nature of the signal control that is being provided. The responses of photoreceptors, horizontal cells, and bipolar cells are mostly graded potentials, which reflect degrees of activation. The activation control mechanism needs to determine and possibly adjust the size of the dark or bright departure from the background average. As illustrated in Figure 6D, the dark and bright channels are registering the differential in luminance of a zone relative to the average surrounding luminance. The horizontal cells are responsible for providing that average. A local zone where the luminance is higher than that average will be signaling bright contrast, and a zone that is below that average will signal dark contrast. These may be fleeting or sustained signals, but either way, the basic mechanism is designed to register the contrast of a figure or figure-related contour that differs from background luminance. The retina needs to registering the amount of luminance departure from background and report the size of that departure, and this seems to call for an analog logic gate.

A formal analog logic gate that might provide this kind of control has been described by Lane Yoder. [71] Yoder's AND-NOT gate registers the difference between X and Y, followed by rectifier clamping of the result to be non-negative. If Z is the output of the AND-NOT gate, then $Z = X - Y$ if $Y <= X$, and is 0 if $Y > X$. These operations can be provided by a differential amplifier, adding a diode that rectifies the signal and clips output if Y is greater than X. Yoder describes how this logic gate can usefully model a number of neural operations, including color classification, smooth-motion control, and memory storage. [71,72] For the present application, the X values would reflect photoreceptor influence on bipolar-cell synapses, with non-inversion of sign at dark synapses and inversion of sign at bright synapses. The Y input would reflect horizontal cell modulation, and the output (Z) would index the degree of stimulus departure from the surrounding luminance baseline, i.e., spatial contrast.

Implemented as a center/surround mechanism, the horizontal cell would register average luminance of the surround zone, and light activation in the center would be scaled as departures from that average. Using the notation cited above, illumination of central cones would provide X, illumination of the surround would provide Y, and the signal being delivered to bipolar cells would be Z. However, the Yoder gate precludes Y being greater than X, so the Z output into dark bipolar cells is clipped (to zero) if the center is brighter than the surround, and input into bright channels is clipped if the center is darker than the surround. As such, the role of the horizontal cells in setting a baseline, such that the activation of the center is reported as a signed and graded departure from average luminance of the zone that surrounds the center.

One might note that such a gating concept provides ample room to adjust for stimulus conditions. One would expect the clip-level to adjust according to overall ambient light level, providing the light adaptation that is traditionally attributed to horizontal cells. If the center/surround zone being registered is very uniform, the value of Y would match X and the clip-value becomes moot. In that case logic gate delivers the center activation to both channels, providing the dual-channel reporting of luminance that was described at the outset. Additionally, the clip level shifts up or down according to image content. If the scene is filled with dark contrast content, the average being registered by the surround will allow a bright stimulus to be especially salient in eliciting a retinal signal. If the scene is filled with bright content, it is the dark stimulus that stands out.

There are numerous findings that call for a gating structure that is not a single center/surround design. As noted previously, bipolar cells have center/surround interactions that are separate from the surround influences on amacrine and ganglion cells. [62] Turner and associates reported that macaque horizontal cells modulate local bipolar-cell response in addition to providing a more general, broad-field surround that can be seen in down-stream neurons. [73] They suggested that the image content can determine the degree to which the bipolar channels are rectified, with high rectification registering localized contrast and low rectification providing a more widespread modulation of activity. (This differential modulation might better be described as clipping adjustments.) Similar findings have been reported in other vertebrates. [74-77]

Selective filtering might provide for differential stimulus encoding. Thoreson & Burkhardt (salamander) found that contrast-response curves of cone-driven bipolar cells varies substantially from cell to cell. [78] And Burkhardt and associates found that the distribution of contrast ranges, which varied from very narrow to quite large, is roughly equivalent to the range of contrasts that can be found in natural images. [29,79] Grabner and associates (ground squirrel) provide a number of alternative physiological mechanisms by which bipolar cells could provide selective encoding of stimulus features. [80]

Some have argued (or implied) that horizontal control is synapse specific, meaning that a differential degree of gating would be provided depending on the exact stimulus conditions being provided to a given bipolar synapse. If this is biologically plausible, this degree of local control would certainly go a long way in registering and signaling the complexities of natural scenes.

### *Perceiving Flicker-fused Letters as Bright or Dark*

I should further articulate how the luminance contrast being reported by the bright and dark channels determines what the observer perceives. If a flicker-fused stimulus conforms to the Talbot-Plateau specifications, it will be judged as being the same brightness as a steady stimulus (see Figs 1 and 2). And across a large range of frequencies, the Talbot-Plateau law also predicts what flicker combinations will produce a stimulus that matches a steady background (see Fig 5). The fewest flicker-fused letters are identified where the law predicts balanced luminance levels. More letters are identified as flash intensities are above or below the balance point, but are they reliably being perceived as brighter or darker than the background?

Figure 8 provides models of bright and dark judgments that respondents made, these being in addition to naming the displayed letters. [56,57] The letters were displayed at 250 Hz, with 50% duty cycle, with judgments being made against three different levels of background luminance -- 4, 8, and 12 $Cd/m^2$. Across the range of flash intensity that was displayed on successive trials, respondents reported whether a perceived letter was bright or dark, or judged the polarity and size of the contrast that was perceived (upper and lower panels in Fig 8, respectively). The green model in each panel specifies the probability that no letter was perceived, this being essentially the inverse of recognition models. The red lines of the upper panels reflect the probability that a bright letter was reported, and the blue lines show the probability that a dark letter was reported. One can see that there is a gradual shift in the relative probability levels, with the cross-over being at or within 10% to the Talbot-Plateau prediction of luminance balance. One can infer that the changes in probability of bright and dark judgments reflects the relative balance of activation of bright and dark retinal channels. The same inference can be made with respect to the scaled judgments of contrast intensity that are shown in the lower panels. The transitions from bright judgments to dark judgments takes place as the luminance of flicker-fused letters crosses the balance point specified by the Talbot-Plateau law. This supports the proposal that luminance contrast is being quantitatively signaled as a departure from the surrounding background, and with horizontal cells control-gating the luminance into bright and dark channels, as describe above.

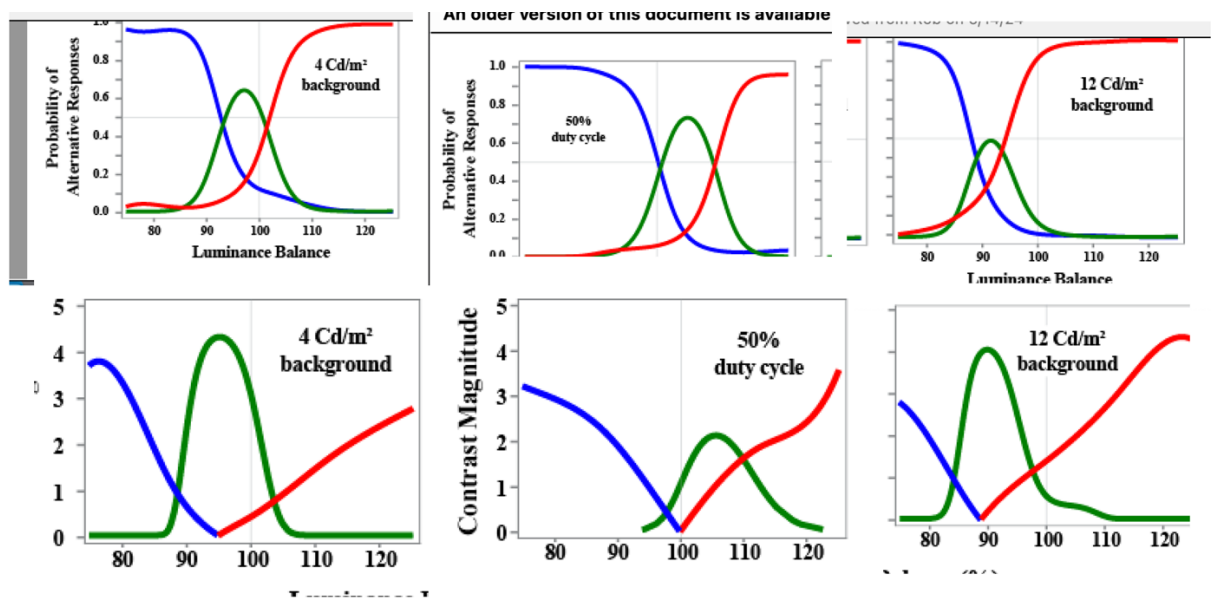

Figure 8. The upper panels display models where respondents reported whether letters were bright (red), dark (blue), or could not be seen (green). The three panels were for judgments made against different levels of background intensity. Lower panels provide corresponding scaled contrast judgments, i.e., judging the level of brightness or darkness of a given letter. The models were derived from two reports that used the same basic research protocols. [56,57]

CODA

The use of flicker-fused stimuli has opened additional ways to evaluate luminance and contrast.  The Talbot-Plateau law provides explicit predictions about what combinations of flash frequency, duration, and intensity will produce a given level of luminance.  The flicker-fused stimulus activates both bright and dark channels, and perceptual judgments reflect the differential degrees of phasic activation.  These channels deliver information that allows perception of image content, which includes the lines and edges that define objects.  Neuromorphic circuits that use bright and dark channels would more closely match biological vision, and would be more effective at discriminating motion, texture, and the contours that define objects.

An extended discussion of retinal mechanisms would have to include amacrine cells, which is beyond the scope of the present discourse.  Here I am only noting the utility of the Talbot-Plateau law, and the benefit of using dark and bright channels as elementary components of neuromorphic vision.

**Acknowledgments**.   Illustration of retinal neurons was done by Lynn Tu.  Funding of research findings was provided by the Neuropsychology Foundation and the Quest for Truth Foundation.